\documentclass{PoS}

\title{Collinear and Transverse Momentum Dependent parton densities obtained with a  Parton Branching Method}

\ShortTitle{Collinear and Transverse Momentum Dependent parton densities obtained with a  Parton Branching Method}

\author{\speaker{Aleksandra Lelek}\\
        Deutsches Elektronen-Synchrotron (DESY)\\
        E-mail: \email{aleksandra.lelek@desy.de}}

\abstract{We present a solution of the DGLAP evolution equations, 
written in terms of Sudakov form factors to describe the branching 
and no-branching probabilities, using a parton branching Monte Carlo method. We 
demonstrate numerically that this method reproduces 
the semi-analytical solutions. 
We show how this method can be used to determine 
Transverse Momentum Dependent (TMD) parton distribution functions, 
in addition to the usual integrated parton distributions functions.
We discuss numerical effects of the  boundary of soft gluon resolution scale parameter on the resulting parton distribution functions.
We show that a very good fit of the integrated TMDs to high precision HERA data can be 
obtained over a large range in $x$ and $Q^2$.}

\FullConference{XXV International Workshop on Deep-Inelastic Scattering and Related Subjects\\
		3-7 April 2017\\
		University of Birmingham, UK}

\begin{document}

\section{Motivation}
Appropriate use of parton distribution functions (PDFs)  (see e.g.~\cite{Collins:2002ey,Collins:2005uv,Nagy:2014oqa}) and a proper treatment of the transverse momentum kinematics in the parton showers (see e.g.~\cite{Collins:2000qd,Hautmann:2012dw,Dooling:2012uw}) are important for comparisons of theoretical predictions with experimental measurements  at the Large Hadron Collider and future colliders experiments. One of the approaches to address these issues is the transverse momentum dependent (TMD) PDFs formalism (see e.g.\cite{Angeles-Martinez:2015sea}). 
The goal of our project is to determine TMD PDFs sets for all flavours, applicable over a broad kinematic region in  $x$, $\mu^{2}$  and $k_t$. We study in particular the effects of the 
soft gluon resolution scale in QCD radiation \cite{shortTMDevolPaper} on the obtained distributions. 
Previous results were reported in \cite{dis16proc}, \cite{ichep16proc}.

\section{Introduction to the method}
\label{sec:1}
We start the discussion from the DGLAP evolution equation for momentum weighted parton densities $xf_a(x,\mu^{2})=\widetilde{f}_a(x,\mu^{2})$
          \begin{equation}\label{eq1}
          \frac{d \widetilde{f}_{a}(x,\mu^{2})}{d\ln \mu ^{2}}=\sum _{b} \int _{x}^{1} \mathrm{d}z \  P_{ab}\left(z,\mu ^{2}\right)\widetilde{f}_{b}\left(\frac{x}{z},\mu ^{2}\right)         
          \end{equation}
          where $a,b$ denote quarks ($n_{f}$ flavours, $n_f\equiv n_f(\mu^2)$) or gluons, $x$ is the longitudinal momentum fraction of the proton carried by a parton $a$, $z$ is the splitting variable and $\mu$ is the evolution variable. The splitting functions $P_{ab}$ have the following structure
          \begin{equation}
P_{ab}\left(z,\mu ^{2}\right)=D_{ab}\left(\mu ^{2}\right)\delta(1-z) + K_{ab}\left(\mu ^{2}\right)\frac{1}{(1-z)_{+}}+R_{ab}\left(z,\mu ^{2}\right)
          \end{equation}
          where $D_{ab}\left(\mu ^{2}\right)=\delta_{ab}d_{a}\left(\mu ^{2}\right)$, $K_{ab}\left(\mu ^{2}\right)=\delta_{ab}k_{a}\left(\mu ^{2}\right)$ and the function $R_{ab}$ does not contain any power divergences (no pieces $\sim(1-z)^{-n}$ for $n=1,2,...$) when $z \rightarrow 1$ but it can contain logarithmic divergences $\ln(1-z)$ which can be however integrated and give finite result. 
     The piece with a delta function and plus prescription corresponds to the virtual and non-resolvable emissions.      
          As long as $P_{ab}\left(z,\mu ^{2}\right)$ has this structure, the formalism  presented here can be applied (LO, NLO, NNLO).
Introducing the soft gluon resolution scale $z_M$ \cite{shortTMDevolPaper},
defining the {\it{Sudakov form factor}} as  
\begin{equation}\label{sud2}
\Delta_{a}(\mu ^{2})=\exp \left(- \int _{\ln\mu _{0}^{2}}^{\ln\mu ^{2}} d\left( \ln \mu ^{\prime 2}\right)\sum_{b}\int _{0}^{z_M} \textrm{d}z zP^{R}_{ba}\left(z,\mu ^{\prime 2}\right)\right)
\end{equation}  
 and using {\it{momentum sum rule}} ($\sum_{c}\int_{0}^{1} \mathrm{d}z zP_{ca}\left(z,\mu^2\right)=0$), Eq.\ref{eq1} can be rewritten
        \begin{equation}\label{eq6}
   \frac{d \widetilde{f}_{a}(x,\mu^{2})}{d\ln \mu ^{2}}=    \sum _{b} \int _{x}^{z_M} dz P^{R}_{ab}\left(z,\mu ^{2}\right)\widetilde{f}_{b}\left(\frac{x}{z},\mu ^{2}\right)   +\widetilde{f}_{a}\left(x,\mu ^{2}\right)  \frac{1}{\Delta_{a}(\mu ^{2})}\frac{d \Delta_{a}(\mu ^{2}) }{d \ln \mu ^{2}} 
         \end{equation}                 
   where $P^{R}_{ab}\left(z,\mu ^{2}\right)=R_{ab}\left(z,\mu ^{2}\right)+K_{ab}\left(\mu ^{2}\right)1/(1-z) $ is the {\it{real}} part of the splitting function.      
 Terms skipped by introducing $z_M$ are of order $\mathcal{O}(1-z_M)$.
   \\
   
Eq.\ref{eq6} has an iterative solution which can be solved by a parton branching method. More details can be found in \cite{shortTMDevolPaper} and \cite{longpaper}.

\section{Collinear PDFs from  parton branching method}
\begin{figure}[h!]
\begin{minipage}{0.5\linewidth}
\centerline{\includegraphics[width=1\linewidth]{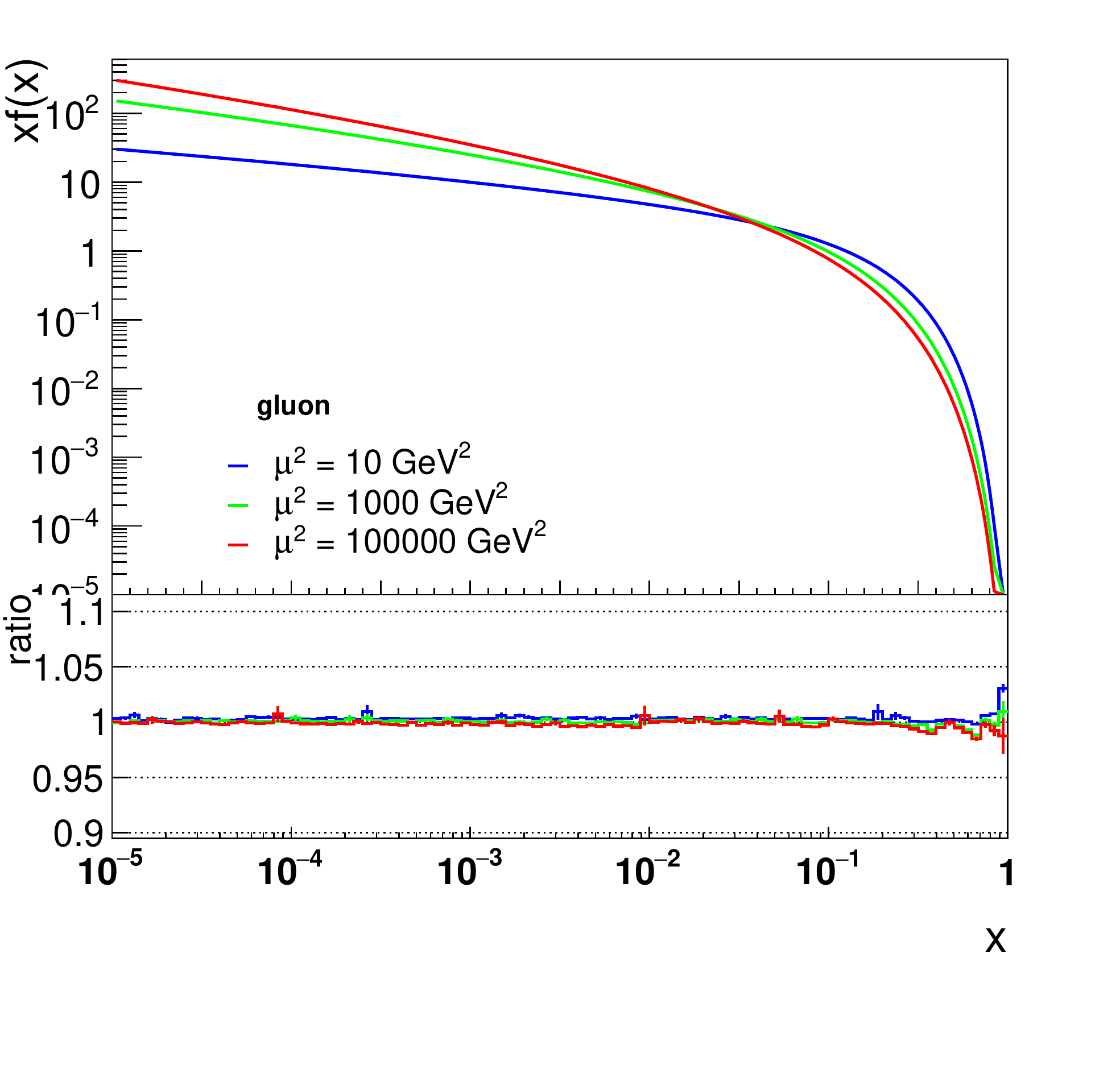}}
\end{minipage}
\hfill
\begin{minipage}{0.5\linewidth}
\centerline{\includegraphics[width=1\linewidth]{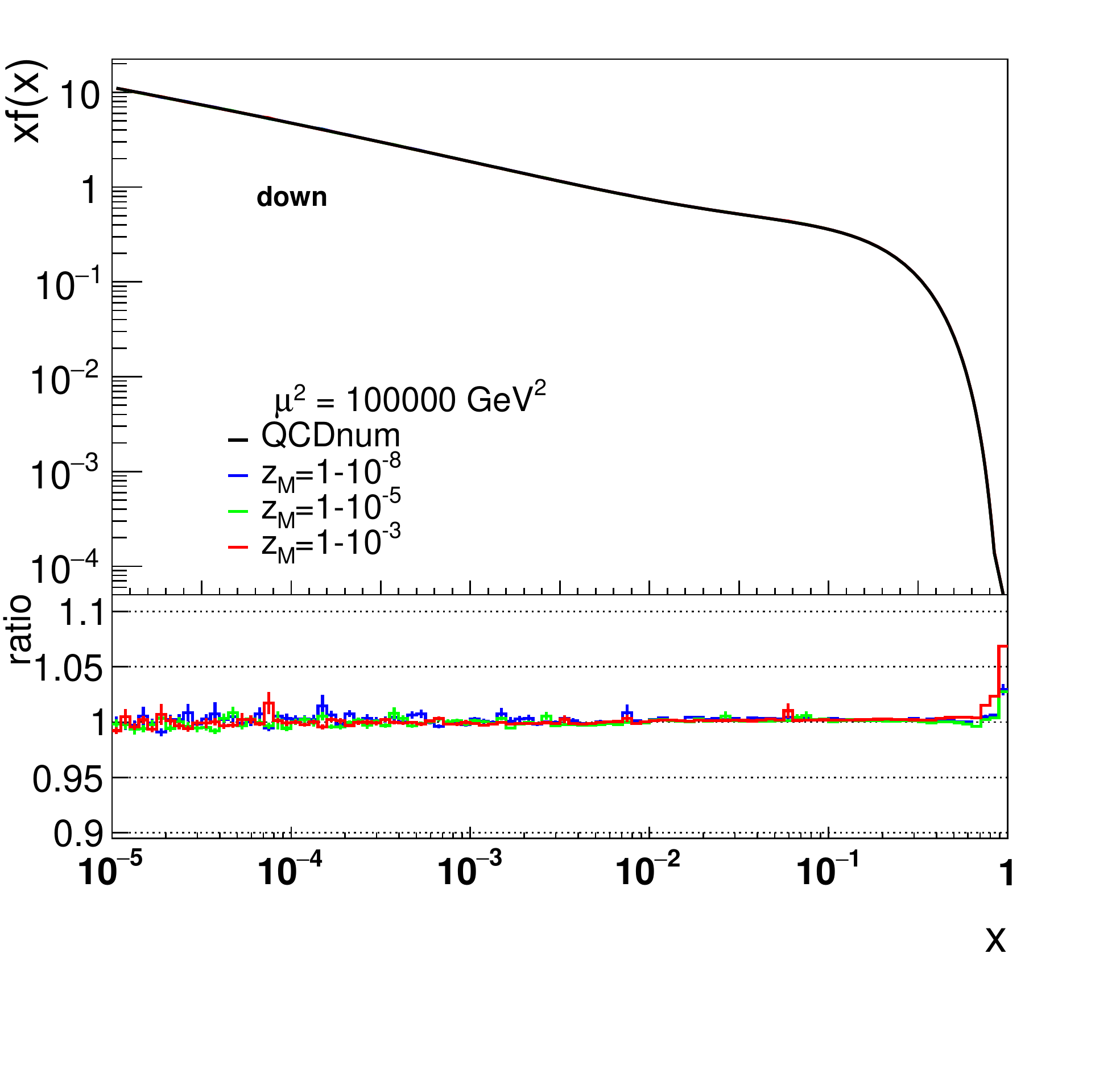}}
\end{minipage}
\hfill
\caption[]{Upper part of the plots: pdfs from the parton branching method for the gluon density at the evolution scale $\mu^2 =10, 10^{3}, 10^{5}\ \textrm{GeV}^{2}$ for $1-z_M=10^{-5}$ (left) and for the down quark density at $\mu^{2}=10^{5}\ \textrm{GeV}^{2}$ for $z_M = 1- 10^{-3}, 1-10^{-5},1-10^{-8}$ (right). Lower part of the plots: ratios of the results from the parton branching method and QCDNUM.}
\label{Fig1:intTMDs}
\end{figure}

In this section  we show the results for collinear PDFs (or integrated TMDs - iTMDs) coming from the parton branching method. In the left hand side of the Fig.\ref{Fig1:intTMDs} we show the results for the gluon density using NLO splitting functions. The evolution is performed with the parton branching method up to the evolution scale $\mu^2 =10, 10^{3}, 10^{5}\ \textrm{GeV}^{2}$ for $z_M=1-10^{-5}$. The obtained distributions are compared with the pdfs calculated from QCDNUM \cite{Botje:2010ay} using the same initial distributions. We illustrate a very good agreement between these two methods. In the right hand side of the Fig.\ref{Fig1:intTMDs} we show the results for the down quark density with the NLO splitting functions for different values of the parameter $z_M = 1- 10^{-3}, 1-10^{-5},1-10^{-8}$ at the evolution scale $\mu^2=10^{5}\  \textrm{GeV}^{2}$. We show that as long as $z_M$ is close to $1$, the results are independent of $z_M$.



\section{TMD  densities}
\begin{figure}
\begin{minipage}{0.27\linewidth}
\vspace{1cm}
\center{\includegraphics[width=0.65\linewidth, trim=0cm 0cm 0cm 0cm, clip ]{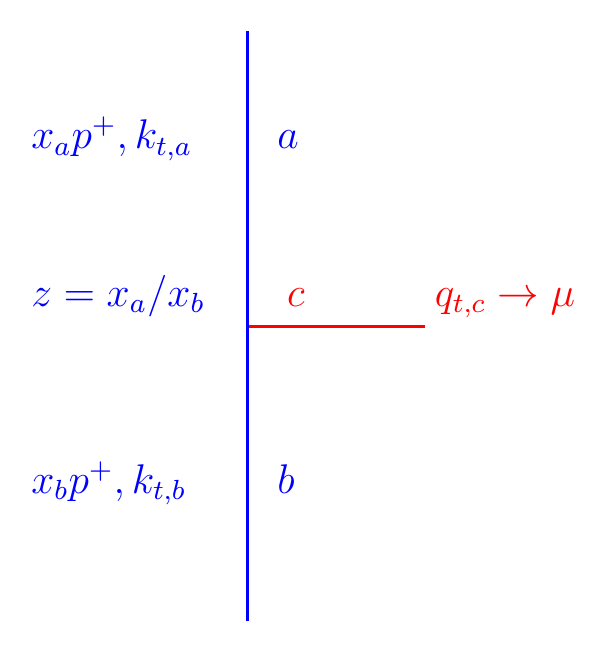}}
\vspace{0.45cm}
\caption{Illustration of a general splitting process: $b\to a +c $.}
\label{Fig:scale-qt-association}
\end{minipage}
\hfill
\begin{minipage}{0.7\linewidth}
\vspace{-2cm}
\includegraphics[width=0.5\linewidth, trim={0cm 0cm 0cm 0cm}, clip ]{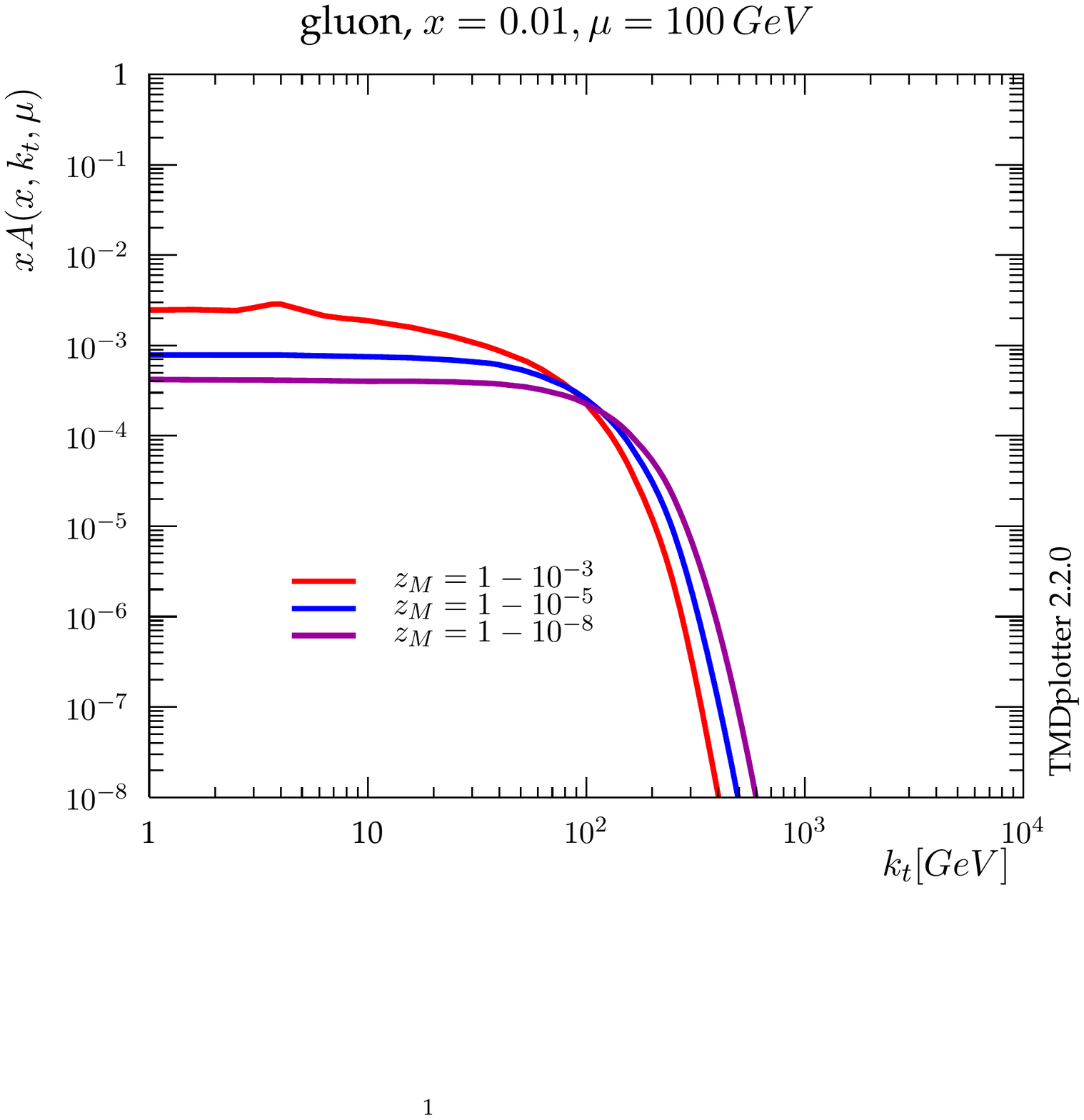}
\includegraphics[width=0.5\linewidth, trim={0cm 0cm 0cm 0cm}, clip ]{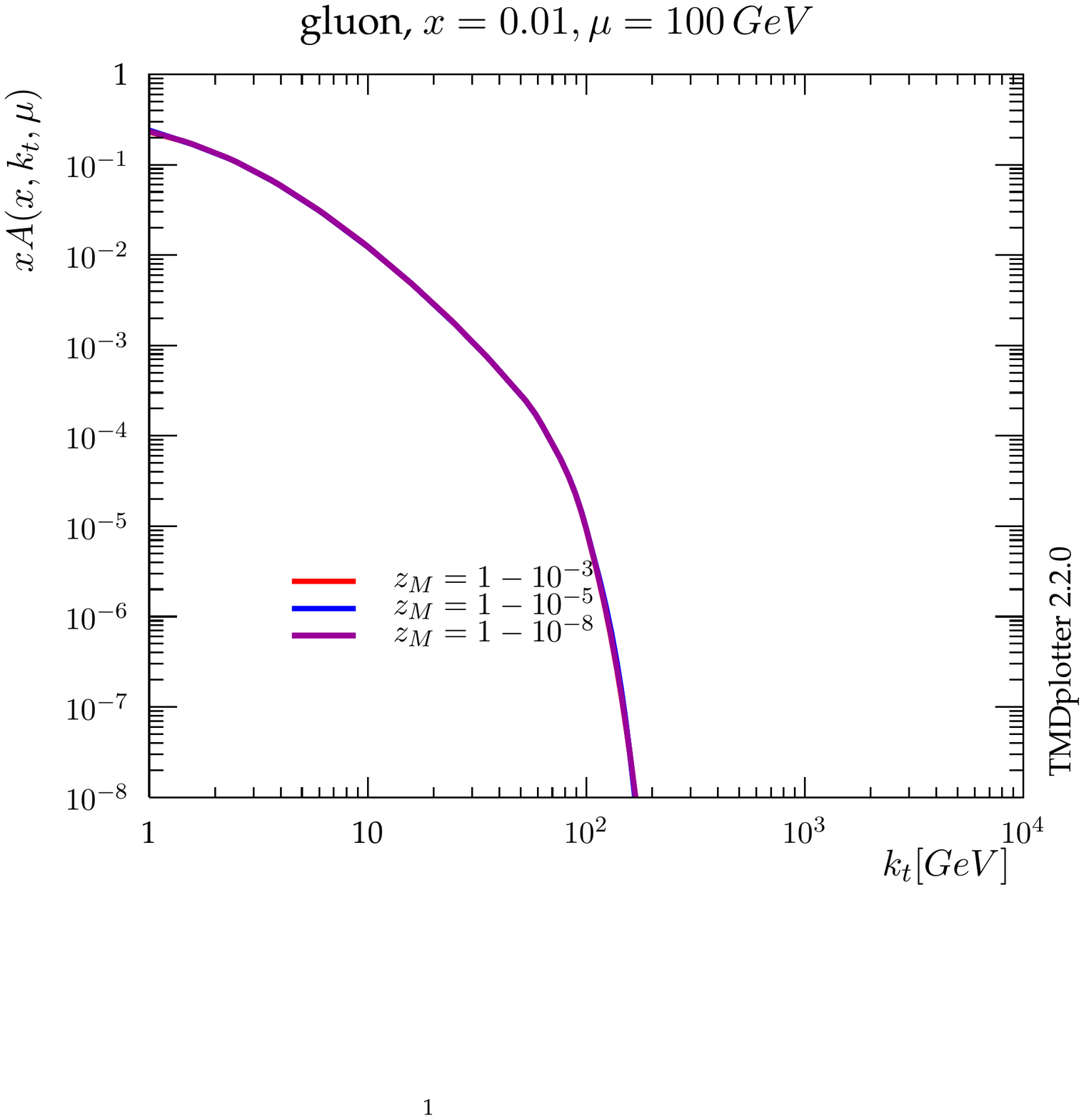}
\vspace{-1.5cm}
\caption[]{TMDs from the parton branching methods for different values of the $z_M$ parameter for $q_t$-ordering (left) and angular ordering (right).}
\label{Fig:TMDs}
\end{minipage}
\hfill
\end{figure}

With the iterative procedure, each resolvable branching is generated and the kinematics in each branching is calculated. Thanks to that, the parton branching method has the great advantage that the transverse momentum $k_t$ of the propagating parton can be calculated and a TMD $ A_a(x,k_t,\mu)$ can be determined.
However, a prescription is needed to relate the evolution variable $\mu$ with the transverse momentum of the emitted ($q_{t,c}$) and propagating ($k_{t,a}$) parton (notation is explained in the Fig.\ref{Fig:scale-qt-association}).
It has been observed \cite{shortTMDevolPaper} that when $q_t$-ordering is used
\begin{equation}
 q_{t,c}^2  =  \mu^ 2,  
  \label{qt-ordering}
\end{equation}
the obtained TMDs depend on the $z_M$ parameter. $z_M$ plays an important role in the large $z$ region- region where soft gluons are emitted. The $z_M$ dependence is a signal that the soft gluons are not treated properly. 
However, if angular ordering condition is used to associate $\mu$ and $q_{t,c}$
\begin{equation}
 q_{t,c}^2  =  (1-z)^2 \mu^ 2,  
  \label{ang-ordering}
\end{equation}
the soft gluons emissions are treated properly - there is no dependence on $z_M$.
This effect is shown in the Fig.\ref{Fig:TMDs}.\\

It is important to stress that the collinear pdf distributions are independent of $z_M$ parameter regardless of the choice of prescription to associate $\mu$ and $q_{t,c}$.  \\


In  the Fig.\ref{Fig2:TMDsflavours} we illustrate that the TMDs for all flavours can be obtained from this method.

\section{Fit to precision DIS data}

The initial parton density distributions have to be determined from a fit of the free parameters to describe inclusive cross sections. We performed the fit within  xFitter package \cite{Alekhin:2014irh} with a method developed in \cite{Hautmann:2013tba,*Jung:2012hy,paperwithSara, paperwithSamatha}.

\begin{figure}[h!]
\begin{minipage}{0.48\linewidth}
\centerline{\includegraphics[width=0.93\linewidth, trim={1cm 3cm 0cm 4.4cm}, clip ]{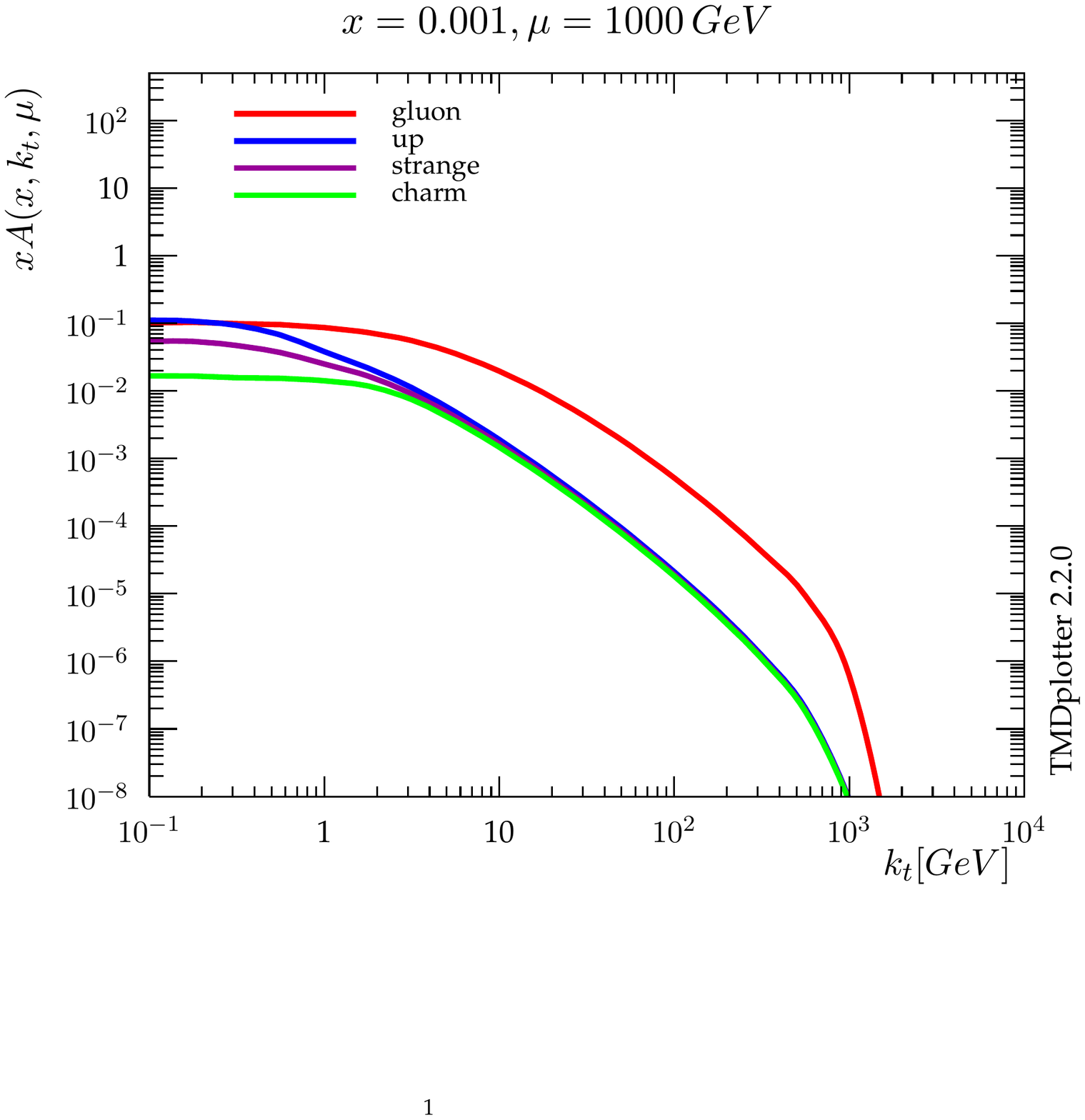}}
\caption[]{TMDs for different flavours from the parton branching method using angular ordering condition from Eq.(\ref{ang-ordering}).}
\label{Fig2:TMDsflavours}
\end{minipage}
\hfill
\begin{minipage}{0.48\linewidth}
\centerline{\includegraphics[width=0.93\linewidth, trim={0cm 0cm 0cm 0cm}, clip ]{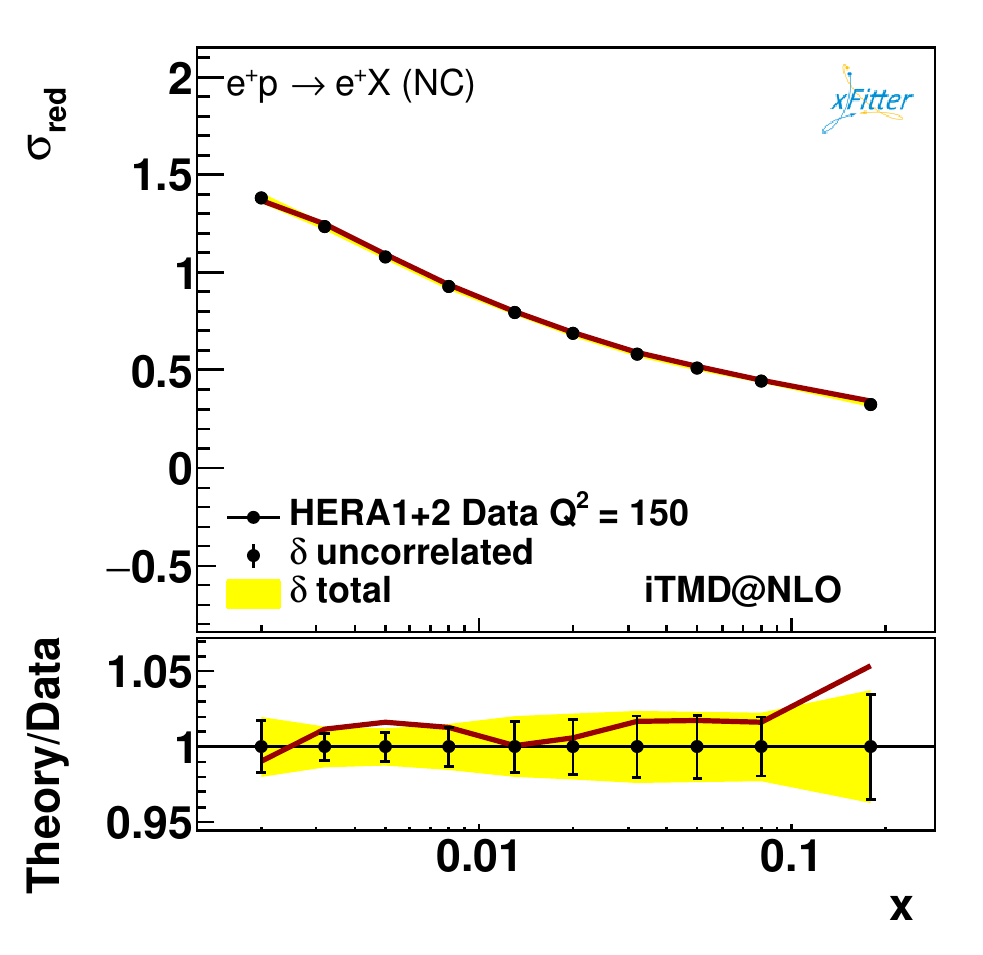}}
\caption[]{Example of a result of the fit of iTMDs from parton branching method to precision DIS data within xFitter.}
\label{Fig2:fitDIS}
\end{minipage}
\hfill
\end{figure}

We used precision measurements in neutral and charged current interactions at various beam energies from HERA 1+2 \cite{Abramowicz:2015mha}  of $\sigma_{red} = d^2 \sigma^{ep}/dx dQ^2 \cdot Q^4 x/(2 \pi \alpha^2 (1 + (1-y)^2))$ (which is the DIS cross section where the photon flux is removed) in the range $3.5< Q^2 < 30000$~GeV$^2$, with $\alpha_s(m_Z) = 0.118$, at a starting scale $\mu_0^2 = 2$~GeV$^2$, masses for heavy quarks as $m_c=1.73 $~GeV, $m_b= 5.0$~GeV and $m_t=175$~GeV together with  a fixed $z_{M}=1-10^{-5}$.
The fit of integrated TMDs  was performed at LO and NLO with experimental uncertainties, for all flavours. A very good $\chi^2/ndf $ was obtained for $3.5< Q^2 < 30000$~GeV$^2$. 
In the Fig.\ref{Fig2:fitDIS} we show an example of the comparison of
the $\sigma_{red}$, obtained from a fit using xFitter,  compared to the precision measurements from HERA.


\section{Summary}
We presented a new approach to solve the DGLAP evolution equation with a parton branching method. We showed that it can reproduce exactly collinear pdfs. We explained how the TMD distributions can be obtained in our method.
We discussed how one could associate the evolution variable $\mu$ with transverse momentum of the emitted and propagating partons to treat properly soft gluon emissions. We showed that with $q_t$-ordering TMD PDFs are not consistently defined 
whereas a consistent set of TMD PDFs can be obtained from the parton branching with angular ordering. 
We determined TMDs with xFitter package from the fit of integrated TMDs to $\sigma_{red}$ precision measurements from HERA 1+2 data at LO and NLO, with experimental uncertainties, for all flavours, in a wide kinematic range of  $x$, $Q^2$ and  $k_t$.

\section{Acknowledgements}
The work presented here was done in collaboration with 
Francesco Hautmann,  
Hannes Jung, 
Voica Radescu and 
Radek ~\v{Z}leb\v{c}\'{i}k.

\end{document}